\documentclass[twocolumn,superscriptaddress,amssymb,amsmath]{revtex4}
\usepackage{graphicx}
\usepackage[english]{babel}

\begin{document}

\title{An Electromagnetic Perpetuum Mobile?}
%\date{\today}
\author{{\O}yvind G. Gr{\o}n}
\affiliation{Oslo University College, Department of engineering,
St.Olavs Pl. 4, 0130 Oslo, Norway}
\affiliation{Institute of Physics, University of Oslo, P.O.Box 1048 Blindern, 0216 Oslo, Norway}
\author{Sigurd K. N{\ae}ss}
\affiliation{Institute of Physics, University of Oslo, P.O.Box 1048 Blindern, 0216 Oslo, Norway}

\begin{abstract}
A charge moving freely in orbit around the Earth radiates according to Larmor’s
formula. If the path is closed, it would constitute a perpetuum mobile. The
solution to this energy paradox is found in an article by C. M. De Witt and B.
De Witt from 1964. The main point is that the equation of motion of a radiating
charge is modified in curved spacetime. In the present article we explain the
physics behind this modification, and use the generalized equation to solve the
perpetuum mobile paradox.
\end{abstract}

\renewcommand{\d}[0]{\partial}
\newcommand{\x}[0]{\mathbf{x}}
\newcommand{\e}[0]{\mathbf{e}}
\newcommand{\myhat}[1]{{\widehat{#1}}}
\newcommand{\basef}[1]{\ul{\omega}^\myhat{#1}}
\newcommand{\conf}[2]{\ul{\Omega}^\myhat{#1}_\myhat{#2}}
\newcommand{\curf}[2]{\ul{R}^\myhat{#1}_\myhat{#2}}
\newcommand{\riemann}[4]{R^\myhat{#1}_{\myhat{#2}\myhat{#3}\myhat{#4}}}
\newcommand{\ricci}[2]{R_{\myhat{#1}\myhat{#2}}}
\newcommand{\ricciu}[2]{R^\myhat{#1}_\myhat{#2}}
\newcommand{\einstein}[2]{G_{\myhat{#1}\myhat{#2}}}
\newcommand{\energy}[2]{T_{\myhat{#1}\myhat{#2}}}
\newcommand{\g}[0]{{\widehat g}}
\newcommand{\fig}[3]{
	\begin{figure}
	\begin{center}
	\includegraphics[width=0.3\textwidth]{#1}
	\caption{#2}
	\label{#3}
	\end{center}
	\end{figure}
}

\maketitle{}

\section{Introduction}
A surprising possibility of a perpetuum mobile seems to exist according to the
usual equations of classical electromagnetism. A situation leading to this
possibility was
described recently by \citet{chiao}. He considered two objects, one neutral and
one charged, orbiting in free fall around the Earth, and wrote:
\begin{quote}
The charged object will gradually spiral in towards the Earth, since it is
undergoing constant centripetal acceleration in uniform circular motion, and
will thereby in principle lose energy due to the emission of electromagnetic
radiation at a rate determined by Larmor's radiation-power formula
[using here SI-units]:
\begin{align}
P_{\textrm{EM}} =& \alpha_q a^2 & \alpha_q =& \frac{q^2}{6\pi\epsilon_0 c^3} \label{prad}
\end{align}
where $P_{\textrm{EM}}$ is the total amount of power emitted in electromagnetic radiation
by the charged object with charge $q$ undergoing centripetal acceleration $a$.
The energy escaping to infinity in the form of electromagnetic radiation emitted by
the orbiting charged object must come from its gravitational potential energy
(which is related by the virial theorem to its kinetic energy), and therefore this
object will gradually spiral inwards towards the surface of the Earth.
\end{quote}

Although everything Chiao says is not only perfectly reasonable, but must also
be correct due to the principle of conservation of energy, the
Lorentz-Abraham-Dirac equation of motion (the LAD-equation) has a strange
consequence as applied to the situation he describes. The covariant form of
this equation is \citep{gron4}:
\begin{align}
f_{\textrm{ext}}^\mu + \Gamma^\mu =& m\dot u^\mu \label{LAD}
\end{align}
where the dot denotes covariant differentiation with respect to the proper
time of the charge, and
\begin{align}
\Gamma^\mu =& \alpha_q\left(\dot a^\mu - \frac{1}{c^2}a^\beta a_\beta u^\mu\right)
\end{align}
is the Abraham four-force, also called the field reaction force. Here $a^\beta$ is
the four-acceleration of the charged body.

The four-acceleration vanishes for geodesic motion. Hence there is no field
reaction force in this case. This means that according to the LAD-equation the
neutral and the charged particle motions are identical, although the
charged particle radiates, and the neutral particle does not.

Since the particles return to the same point of the path after one period, they
can continue this motion for an unlimited amount of time. And the charged
particle would continue radiating all the time. Hence, this is a perfect
perpetuum mobile. One gets radiation energy for nothing from a system working
cyclically.

Obviously this in conflict with the principle of conservation of energy, so
something must be wrong. There are two equations involved in this energy
paradox, the Larmor equation and the LAD-equation.

One solution of this energy paradox would be that a charge falling freely does
not radiate. One could argue for this by noting that a freely moving particle
defines a local inertial reference frame, i.e. a frame in which Newton's 1st
law is valid. Let the neutral object mentioned by Chiao be an observer. As
measured by this observer the charge is at rest in an inertial frame, and hence
it does not radiate. However, a thorough analysis proves that this is not
the solution to the paradox \citep{boulware,fugmann,fugmann2,hirayama,gron5}.
It has been shown that, as measured by an observer that is not falling
freely, a freely falling charge radiates with a power given by Larmor's
formula, and it extension to a non-inertial reference frame. No generalization
to curved spacetime has been made, but since the above result shows that
it does not matter for the radiated effect whether a charge is accelerated by
gravitational or normal forces in flat spacetime, we will assume that this
also holds, at least in the Newtonian limit of weak field and low relative
velocities, in curved spacetime.

The other equation is the equation of motion of the charged, radiating
particle, the LAD-equation. Maybe there is something wrong with this. Could
there be a non-vanishing field reaction force on a freely falling charge? An
argument against this possibility is the following. According to the principle
of equivalence no local measurement should be able to distinguish between being
at rest in flat space and being freely falling in curved space. As observed by
the comoving neutral object there is no force acting upon the charged particle.
Hence it should remain at rest relative to the neutral object, and this is a
reference independent result.

This argument has, however, a weak point. The principle of equivalence has a
local character. The mentioned equivalence is only valid as far as the
measurements does not reveal a possible curvature of space. So, if there is a
non-local interaction between the charge and the curvature of spacetime, due to
the non-local character of its electromagnetic field, this may modify the
equation of motion of the charge. The necessity and nature of such a
modification will be explained in the next sections.

\section{The field reaction force in curved space}
Before we can understand the solution to the energy paradox introduced above,
we need to know the correct form of the equation of motion of a charged
particle in curved space. It was, in fact, deduced more than forty
years ago \citep{brehme,hobbs}, but is not well known outside a rather small group of
physicists having performed research within this field. The reason it
not that this knowledge is not important - the equation of motion of
a charged particle in curved space is one of the fundamental equations
in classical electrodynamics - but that the equation is mathematically
rather complicated. Nevertheless, the physics behind the equation can
be understood, as we shall explain in this article.
\subsection{An introduction to the electromagnetic self force problem in curved spacetime}
Let us consider the motion of a classical charged point particle due to
the electromagnetic force. The path of the particle is $\mathbf
x'(\tau)$, and we shall also need a label for a general point in
space-time, which will be $\mathbf x$. To make the notation compact,
it is usual to associate the prime also with components of vectors
taken at a given point. Thus, $j^\mu$ is understood to be $j^\mu(\mathbf x)$,
while $j^{\mu'} = j^{\mu'}(\mathbf x')$.

In general, the force on a body due to the electromagnetic interaction
is given by the Lorentz equation
\begin{align}
\frac{df^\mu}{dV} =& F^\mu{}_\nu j^\nu
\end{align}
where $\frac{df^\mu}{dV}$ is the force per unit volume on
the body, the tensors $F_{\mu\nu} = A_{\nu,\mu}-A_{\mu,\nu}$ and
$j^\mu$ are the electromagnetic field and current density, respectively,
and $A^\mu$ is the electromagnetic potential. The force on the whole
body, then, is
\begin{align}
f^\mu =& \int_V F^\mu{}_\nu j^\nu dV
\end{align}
In the case of a point particle with charge $q$ four-velocity
$\mathbf u$ and current density $j^\mu(\mathbf x) = q\int u^\mu
\delta(\mathbf x,\mathbf x'(\tau))d\tau$, where $\tau$ is
the proper time of the particle, the Lorentz equation
reduces to
\begin{align}
f^\mu =& q F^\mu{}_\nu u^\mu
\end{align}
The $\delta(\mathbf x,\mathbf x')\equiv g^{-\frac 1 2}\delta(
\mathbf x-\mathbf x')$ used in the expression for the
four-current is the invariant delta-function,
which is defined such that any volume integral over $\mathbf x$
is 1 if it contains $\mathbf x'$, and 0 otherwise. furthermore
$g$ is the absolute value of the determinant of the metric.

Since the particle is charged, it not only reacts to $F_{\mu\nu}$, but
also acts as a source of it, according to the electromagnetic wave
function
\begin{align}
A^{\mu;\alpha}{}_\alpha - R^\mu_\alpha A^\alpha =& -\mu_0 j^\mu
\label{EMWave}
\end{align}
with $R_{\mu\nu}$ being the Ricci curvature tensor. The solution
of this equation depends on the motion of the particle, encoded in
$j^\mu$, which is again determined by the Lorentz equation, so
these two equations are coupled, the coupling being the particle's
reaction to its own field.

\subsection{Green functions}
Since these equations are linear, it is possible to split $j^\mu$
into points, solve the equation for each point, and then take the
sum of the results. Furthermore, we do not need to know the
amplitude of $j^\mu$ at each of these points when we solve the
equation, as the linearity makes it possible to multiply by this
afterwards. This way, much of the work in solving
the equation can be done without involving $j^\mu$ itself.
This is called the Green function approach to solving
differential equations.

In general, given some linear differential equation
\begin{align}
L(\phi(\x)) =& f(\x)
\end{align}
where $L$ is a general linear differential operator (for example $L =
\frac{\d^2}{dx^2} + y\frac{\d}{\d y}$), a Green function defined
by
\begin{align}
L(G(\x,\x')) =& \delta(\x-\x') \label{green1}
\end{align}
where $\delta(\x-\x')$ is the Dirac delta function,
solves the original differential equation through
\begin{align}
\phi(\x) =& \int G(\x,\x') f(\x') d\x' + S
\end{align}
where $S$ is a surface term that can be taken to be zero
when the integral goes over all of spacetime.

In many cases, it is easier to find the Green function than
the unknown function, and so it is mainly used for convenience, but
it also serves to make the causal connections inherent in the
differential equation explicit. From the definition, we see that
the $G(\x,\x')$ is the effect on the spacetime point $\x$ of
the point $\x'$, and thus can be seen as the degree to which
the two points are causally connected (as far as the differential
equation is concerned). For example, in flat, 3+1-dimensional
spacetime, electromagnetic radiation moves only on the light
cone, and the Green function for electromagnetic radiation
is therefore zero everywhere outside of the light cone, so
in this case $G \propto \delta(\sigma)$. The $\sigma$ appearing
here is commonly used in the theory of geodesics, and
is defined as $\sigma \equiv \frac 1 2 s^2$, with $s$ being
the geodetic interval between two points ($\mathbf x$ and
$\mathbf x'$ in this case). It is negative for a time-like
interval, positive for a space-like interval, and zero for
a light-like interval.

The Green function defined above is a scalar quantity. However,
the equation we are considering, the electromagnetic wave equation,
is a vector equation.
A straightforward generalization of
equation (\ref{green1}) for this case is
\begin{align}
{L^\mu}_\nu({G^\nu}_{\mu'}(\x,\x')) =& {\delta^\mu}_{\mu'}\delta(\x-\x') \\
A^\mu =& \int {G^\mu}_{\mu'} f^{\mu'} d\x' \label{green2}
\end{align}
This equation is correct\footnote{In equation (\ref{green2}), indices
belonging to different points
are being compared directly in the Kronecker delta ${\delta^\mu}_{\mu'}$.
Since basis vectors in general may vary from point to point,
the meaning of the components is position dependent, and one should
therefore be careful when directly comparing them. A generalization of
the Kronecker delta $\delta^\mu{}_{\mu'}$ called the parallel propagator
$g^\mu{}_{\mu'}$ takes care of this. In principle, then, we should replace
the Kronecker delta with the parallel propagator in equation (\ref{green2}).
However, since the delta function constrains the point $\mathbf x'$ to
coincide with $\mathbf x$, this complication does not occur here, and
we can keep the Kronecker delta.} as it stands, but to make it an invariant
expression, we need to replace the delta function with its invariant
version $\delta(\mathbf x,\mathbf x')$, and the integral needs a
volume element of $\sqrt{g}$. Rectifying this, we arrive at
\begin{align}
{L^\mu}_\nu({G^\nu}_{\mu'}(\x,\x')) =& {\delta^\mu}_{\mu'}\delta(\x,\x') \\
\phi^\mu =& \int {G^\mu}_{\mu'} f^{\mu'} \sqrt{g} d\x'
\end{align}
Applying this to the electromagnetic wave equation, we get
\begin{align}
\Box {G^\mu}_{\mu'} - {R^\mu}_\nu {G^\nu}_{\mu'} =& -\mu_0 {\delta^\mu}_{\mu'}\delta(\x,\x') \label{EMGreen}\\
A^\mu =& \int {G^\mu}_{\mu'} j^{\mu'} \sqrt{g} d\x' \label{EMGreen2}
\end{align}
where the constant factor $-\mu_0$ has been absorbed into $G$
due to convention.

\subsection{Generalized LAD equation for curved space}
Equation (\ref{EMGreen}) differs from equation (\ref{EMWave})
in one important aspect: It is independent of $j^{\mu'}$, and
thus not coupled to the motion of the particle. This is the main
reason for introducing Green functions. It may still
be difficult to solve, but it can be solved once and for all,
and then applied to specific cases with equation (\ref{EMGreen2}).

Let us assume that we already know the Green function for the
wave equation; we can then use it to find the electromagnetic potential:
\begin{align}
A^\mu(x) =& \int G^\mu{}_{\alpha'}(x,x')j^{\alpha'}(x')\sqrt{g'}d^4x' \notag \\
=& q\int G^\mu{}_{\alpha'}(\mathbf x,\mathbf x'(\tau)) u^{\alpha'} d\tau
\label{AG}
\end{align}
Since this is an explicit expression for $A^\mu$, we can insert it into
the Lorentz equation, resulting in
\begin{align}
f^\mu(\tau) =& q^2
 \int d\tau'
 \big(G_{\rho\alpha'}{}^{;\mu}(\mathbf x'(\tau),\mathbf x'(\tau'))\notag\\
 &-G^\mu{}_{\alpha';\rho}(\mathbf x'(\tau),\mathbf x'(\tau'))\big)
 u^{\alpha'}(\tau')u^\rho(\tau) \label{fullint}
\end{align}
Here, the general point $\mathbf x$ has been replaced by the position
of the particle, $\mathbf x'$, as that is the point where we use
the Lorentz equation. The physical interpretation of this equation
is that each point $\mathbf x'(\tau')$ on the world line of
the particle gives a contribution to the force acting on some
chosen point $\mathbf x'(\tau)$, the effect of $\mathbf x'(\tau')$
on the potential, and therefore on the force, being given by
the Green function $G^\mu{}_{\mu'}(\mathbf x'(\tau),\mathbf x'(\tau'))$.

The case $\tau = \tau'$, where
emission of and reaction to the field happens at the same time,
stands out as both promising due to its locality, and daunting due
to the fact that the infinite charge density of a point charge sets up an
infinitely strong field at its location.

The contribution from this case was computed by \citet{brehme} in 1960
and \citet{hobbs} in 1968 as
\begin{align}
\Gamma^\mu-\frac 1 2\alpha_q\left(R^\mu{}_\nu u^\nu c^2
	+  R_{\alpha\beta}u^\alpha u^\beta u^\nu\right)
\end{align}
here $\Gamma^\mu$ is the Abraham four-force. The part of the integral
(\ref{fullint})
where the ``absorption'' time $\tau$ is
earlier than the emission time $\tau'$, can be eliminated on
grounds of causality, and we are left with
\begin{align}
f^\mu(\tau) =& \Gamma^\mu - \frac{1}{2}\alpha_q\left(R^\mu{}_\nu u^\nu c^2
	+  R_{\alpha\beta}u^\alpha u^\beta u^\nu\right) \notag \\
&+q^2\int_{-\infty}^{\tau^-} d\tau'
 \left(G_{\rho\alpha'}{}^{;\mu}- G^\mu{}_{\alpha';\rho}\right)
 u^{\alpha'}(\tau')u^\rho(\tau) \label{GLAD}
\end{align}
This is a generalization of the LAD equation we used earlier,
which stops after the first term on
the right hand side, and is therefore a local equation.
It is on this generalized equation that \citet{dewitt2} base their
calculations.

Though we have seen how the integral term, usually called the tail term,
originates analytically, it is still not obvious that it can be anything
but zero. We will therefore take another detour before returning to
the DeWitts' results, in order to show how this can be.

As we remember from the derivation of the term, it represents
the force from a field that is generated at the point
$\mathbf x'(\tau')$, and later reaches the point $\mathbf x'(\tau)$.
The behavior of the electromagnetic field is encoded in
$G$, but before calculating, we already have some expectations
about its form from our experience with electromagnetism:
\begin{enumerate}
\item Electromagnetic fields in vacuum should move at the
speed of light, $c$.
\item It should be impossible for a particle to catch up with
a light-like field it has emitted.
\end{enumerate}
From these two points, we would conclude that the only possible
way a particle can interact with itself is when emission and
absorption of the field happens at the same point, so there
should be no nonlocal contributions, and therefore no tail
term!

Our intuition turns out to be wrong on both these points, and
both of them can give contributions to the tail term.

\section{The physics behind the non-local field reaction in curved space-time}
\subsection{Overtaking light}
The latter premise for the assertion that there should be
no tail term, namely that it should be impossible for a particle
to catch up with a light-like field it has emitted, is the easiest
to prove wrong, as gravitational lensing provides a simple
counterexample.

As an extreme case of gravitational lensing, consider a particle
suspended at a constant Schwarzschild
coordinate of $r = \frac 3 2 R_S$, i.e. at the innermost stable
orbit. It then emits a photon perpendicularly to the direction
towards the black hole. The photon will then reencounter the
particle after one orbit. The particle has, effectively, caught
up with the photon by taking another path through space-time.

Less extreme examples of lensing also make the same effect possible,
though a relative velocity between the mass and the particle will
be necessary in those cases.

\subsection{Slow light}
In the above case, it was possible to catch up to light because light
took another path through spacetime, though it still moved at the speed of
light. It is more difficult to see how light can end up moving more slowly
than $c$ through empty space. However, this turns out to be the most
important contribution to the tail term.

To see how the wave equation can lead to parts of the field moving
more slowly than the speed of light, and thus causing dispersion, so
that
an initially sharp pulse develops a trailing tail, let
us first consider the spherically symmetric case in flat
spacetime, and vary the number of spatial dimensions. These
cases provide one of the conceptually simplest examples
of dispersion, and will turn out to be mathematically related
to cases in curved, 3+1-dimensional spacetime.

In flat spacetime, it is possible to decompose the wave equation,
which is a vector equation, into one scalar wave equation for
each dimension of space-time, with all the scalar equations
being independent. These scalar equations have the form
\begin{align}
\phi^{;\mu}{}_\mu = -\rho \label{scalar}
\end{align}
and solving this equation is equivalent to solving the vector
equation. In curved spacetime, this is no longer the case,
but the solutions to the vector and scalar equations generally display
similar behavior.

In polar coordinates with spherical symmetry, equation (\ref{scalar})
takes the form $\left(-\frac{1}{c^2}\frac{\d^2}{\d t^2} + \frac{\d^2}{\d r^2}
+\frac{d-1}{r}\frac{\d}{\d r}\right)\phi = -\rho_r$, with $\rho_r$
being the angular integral of $\rho$, and $d$ being the number of spatial
dimensions. The equation for the Green function corresponding to this
is
\begin{align}
\left(-\frac{1}{c^2}\frac{\d^2}{\d t^2}+\frac{\d^2}{\d r^2}+\frac{d-1}{r}\frac{\d}{\d r}
\right)G =& -2c\delta(t)\delta(r) \label{ddimrad}
\end{align}
Note that the only thing separating the cases with different number
of dimensions, is the factor in front of the $\frac{\d}{\d r}$
term, but this factor is critical for its behavior. It possesses
simple solutions for the cases $d=1$ and $d=3$, but is complicated to
solve for other cases, especially the even ones. \citet{mphuyg}
discusses this thoroughly in an article on the Huygens' Principle.
He only considers the homogeneous version of the equation,
\begin{align}
\left(-\frac{1}{c^2}\frac{\d^2}{\d t^2}+\frac{\d^2}{\d r^2}+\frac{d-1}{r}\frac{\d}{\d r}
\right)\phi =& 0 \label{ddimhom}
\end{align}
but since the right hand side of equation (\ref{ddimrad}) is given by
delta functions, they only differ at the initial moment, and so
the solution to equation (\ref{ddimrad}) can be found by solving
equation (\ref{ddimhom}) with an appropriate initial condition.

To see that the case for 1 spatial dimension and our familiar 3 spatial
dimensions are exceptionally simple, Brown suggests that one makes a
change of variable by introducing
$\psi = r^\frac{d-1}{2}\phi$ (our notation switches the role of $\phi$
and $\psi$ compared to his), which leads to the differential equation
\begin{align}
\left(-\frac{1}{c^2}\frac{\d^2}{\d t^2}-\frac{\d^2}{\d r^2}+\frac{(d-1)(d-3)}{4r^2}\frac{\d}{\d r}
\right)\psi =& 0
\end{align}
The first order term vanishes for $d=1$ and $d=3$, giving the simple
equation $\frac{1}{c^2}\frac{\d^2\psi}{\d t^2} = \frac{\d^2\psi}{\d r^2}$,
which can be rewritten as $\left(\frac{d}{dr}-\frac{1}{c}\frac{d}{dt}\right)
\left(\frac{d}{dr}+\frac{1}{c}\frac{d}{dt}\right)\psi = 0$, and so obviously
has solutions
\begin{align}
\psi &= f(r-ct)+g(r+ct)
\end{align}
for any function $f$ and $g$\footnote{Since $r$ is a radial coordinate we also
have a mirror boundary condition or $\frac{\d\psi}{\d r} = 0$ at $r=0$.
Due to the linearity of the equation, this is equivalent to extending the
domain of $r$ to $(-\infty,\infty)$, solving the equation, and then finding
the real solution as $\psi_{r\in(0,\infty)}(t,r) = \psi_{r\in(-\infty,\infty)}
(t,r) + \psi_{r\in(-\infty,\infty)}(t,-r)$. The boundary condition does
not change the point that the solution moves at a single, well-defined speed, $c$.}
and then It is easy to see that this solution consists
of signals moving at a single, well-defined speed: $c$.
This exceptional case ($d=3$) is what our
intuition for the wave equation, and thus behavior of light, is based
on, and the behavior of electromagnetic fields in other numbers of dimensions
therefore end up being counterintuitive. For a general $d$, we cannot
expect the first order term to cancel, and therefore do not end up with
an an equation that leads to this kind of simple propagating solution.

Brown goes on to consider the case $d=2$, and by assuming separability,
$\phi(r,t)=f(r)g(t)$, rewrites equation (\ref{ddimhom}) as
\begin{align}
r^2\frac{d^2f}{dr^2}+r\frac{df}{dr}+r^2f =& 0 &
\frac{d^2g}{dt^2} =& -c^2g
\end{align}
The equation for $f$ is the zeroth order Bessel equation, and is
solved by the corresponding zeroth order Bessel functions $J_0(r)$
and $Y_0(r)$. By using the integral representation for $J_0$,
$J_0(r) = \frac 2\pi\int_0^\infty \sin\left(\cosh(\theta)r\right)d\theta$,
and a solution $g(t) = sin(ct)$ of the equation for $g$, Brown
finds a solution for $\phi$ to be
\begin{align}
\phi =& \frac 1\pi\int_0^\infty\left[\cos\left(\cosh(\theta)r-ct\right)
-\cos\left(\cosh(\theta)r+ct\right)\right]d\theta
\end{align}
This, he points out, is a superposition of waves traveling with
speeds of $v = \frac{c}{\cosh(\theta)}$, which takes all values
from 0 to $c$. So in this case, light moves at the speed of light,
as well as \emph{all lower velocities}!

To see how this applies to the actual Green equation, which differs from
the equation considered by Brown by being inhomogeneous, we present
numerical solutions for the cases of
2, 3 and 4 spatial dimensions, and with delta functions
approximated by Gaussians. The result can be seen in the figures \ref{ndim2}-\ref{ndim4}.

\fig{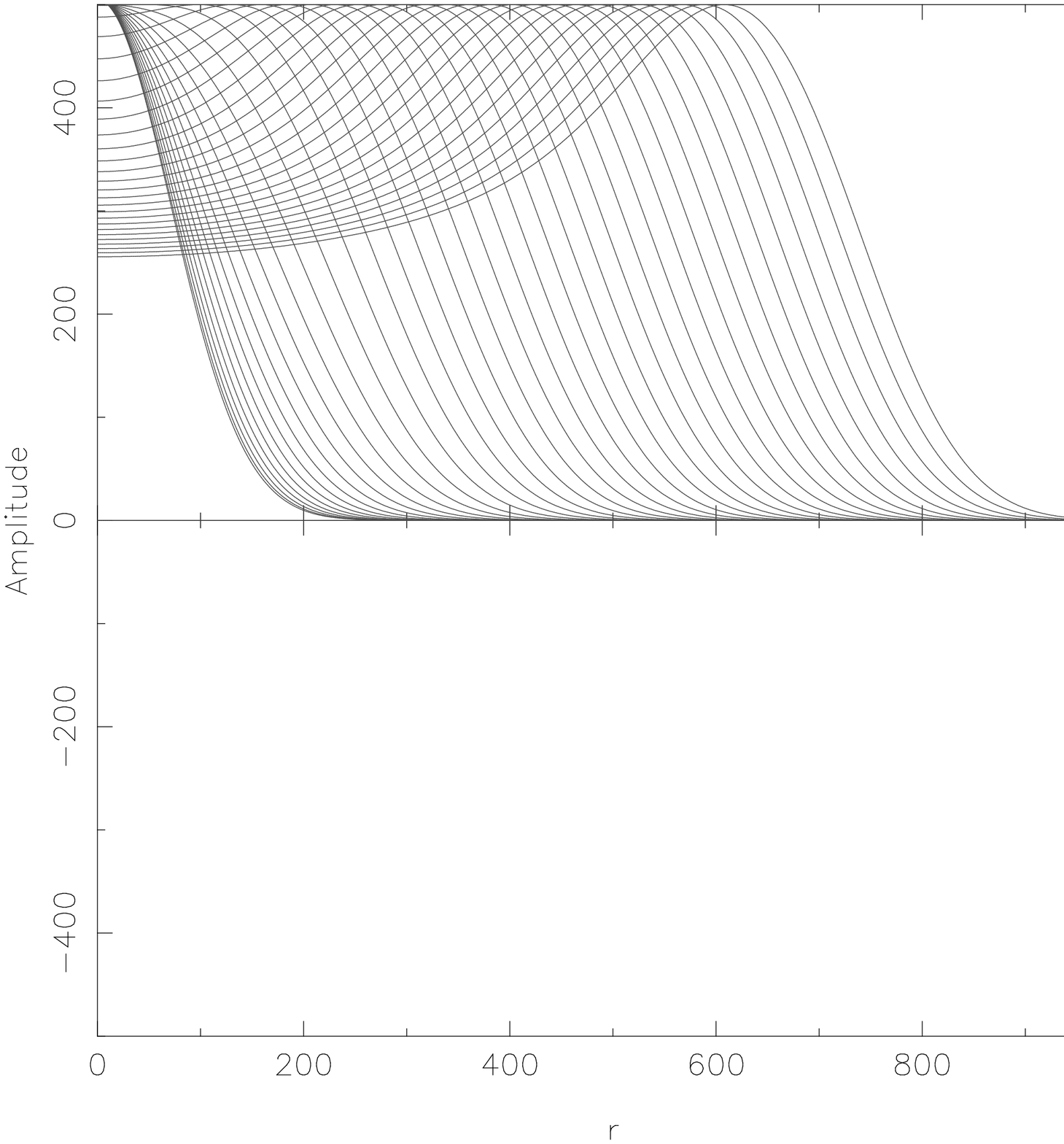}{Time series for the solution of the equation
$\left(-\frac{1}{c^2}\d_t^2+\d_r^2+\frac{d-1}{r}\d_r\right)G = -2c\delta(t)\delta(r)$
with $d = 2$ and Gaussians instead of delta functions, and $c=1$.
$r$ runs along the horizontal axis, and $G$ along
the vertical one.
This is a set of superimposed curves, one for each time step.
The general behavior is that of a pulse propagating to the right,
but the value behind the pulse does not reach 0, making the final
result a lopsided Gaussian. Since we have spherical symmetry, the time
series in this figure corresponds to an initial central peak evolving
into a spreading circular ``hill'', with a nonzero ``elevation'' inside,
which is the ``tail''.}{ndim2}
\fig{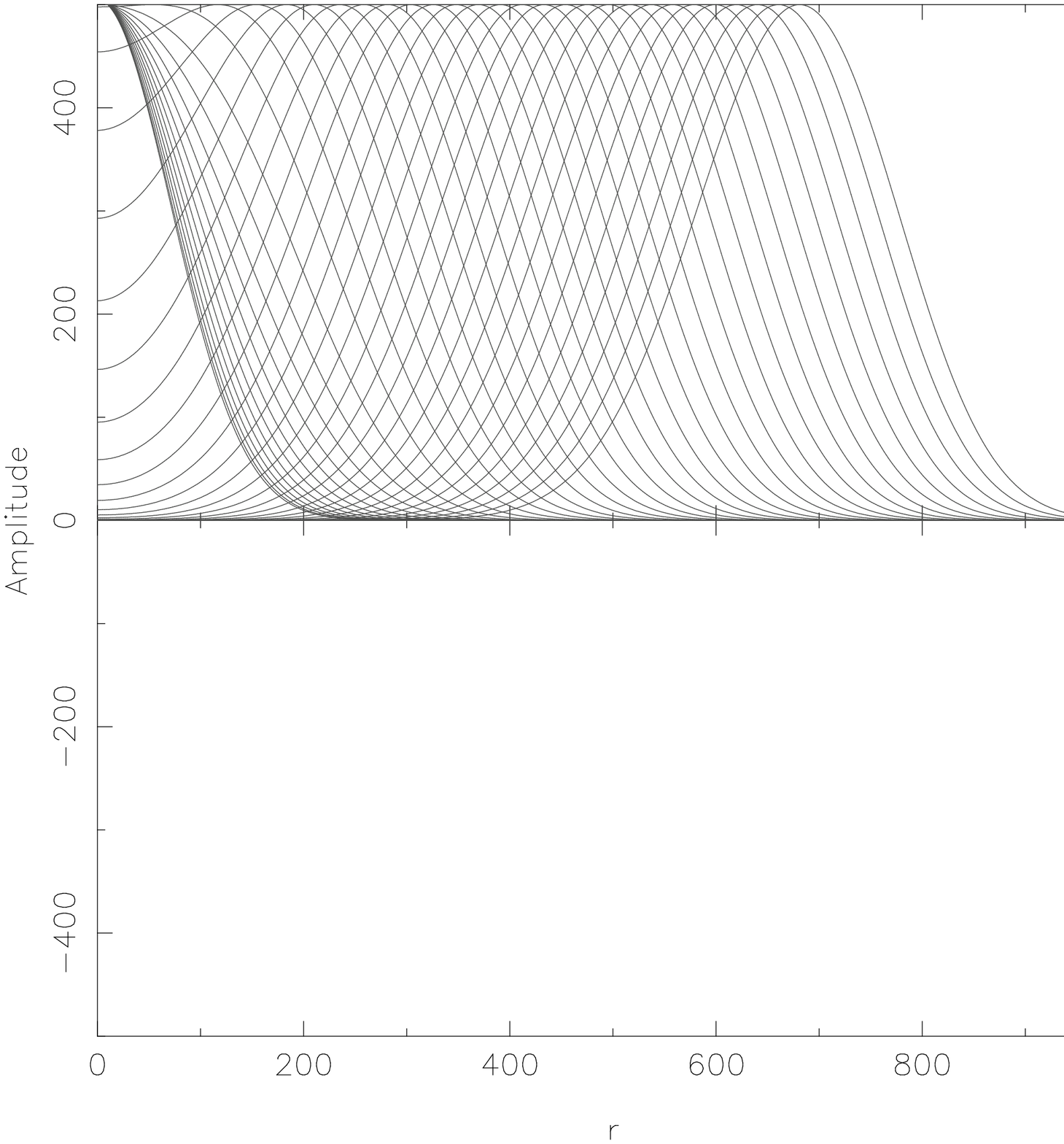}{As figure \ref{ndim2}, but for $d=3$. This time
the value behind the pulse does reach 0, and we get a
normal propagating Gaussian, corresponding to a propagating delta
delta function. Since the amplitude is zero before and after the
wave front, radiation propagates only on the light cone in this case.}
{ndim3}
\fig{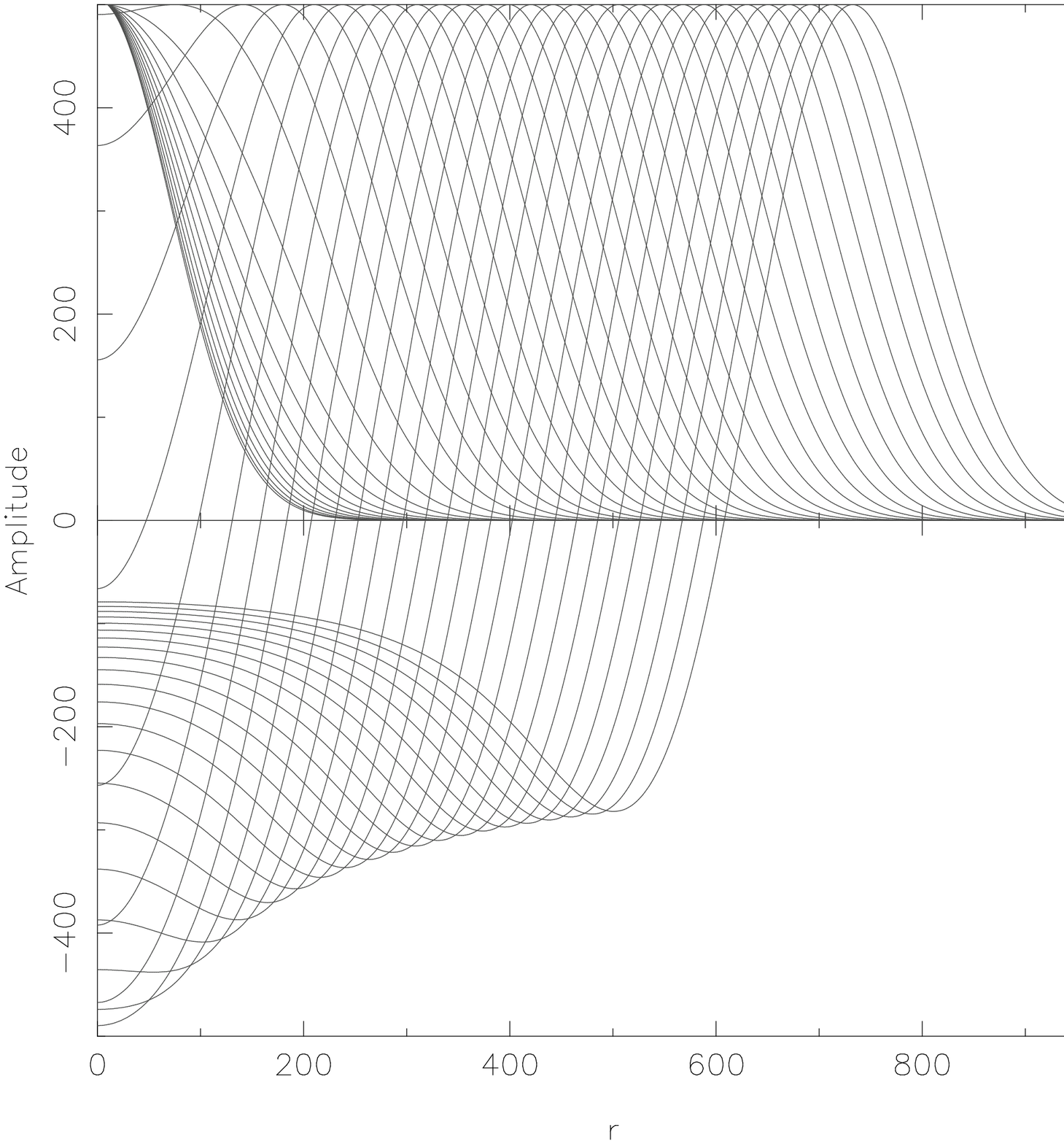}{As figure \ref{ndim2}, but for $d=4$. This bears
more resemblance to the case of $n=2$ than $n=3$, in that the
amplitude behind the wave front does not reach zero, but this time
it is due to it falling too rapidly, overshooting zero.}{ndim4}

From these solutions, we see that the factor in front of the
$\frac{\d}{\d r}$ term of equation (\ref{ddimrad})
controls how fast the Green function
falls after the wave front has passed. For $d=2$, which
gives a factor of $\frac 1 r$, the fall after the wave front
is less than the rise that preceded it, so the function
develops a tail. For $d=4$, with a factor of $\frac 3 r$,
the fall is too great, and the solution overshoots zero,
again leaving a tail. The case $d=3$ has the exact balancing
between fall and rise needed to avoid a tail. We again see that
not having a tail is not a general feature of the wave equation,
but a special case.

To see how this applies to curved 3+1-dimensional spacetime,
we now consider the form of the spherically symmetric scalar
wave equation in the Schwarzschild spacetime:
\begin{align}
\left(-\frac{\beta}{c^2}\frac{\d^2}{\d t^2}+\frac{1}{\beta}
\frac{\d^2}{\d r^2}
+\frac{1+\beta}{r}\frac{\d}{\d r}\right)G =& -2c\delta(t)\delta(r) \label{srgwe1}
\end{align}
with the shorthand notation $\beta = 1-\frac{R_S}{r}$.
This equation is expressed in terms of Schwarzschild radial coordinates,
which behave as flat-space polar coordinates only in the limit $r
\rightarrow \infty$. This prevents us from directly comparing it to
equation (\ref{ddimrad}). To remedy this, one can introduce orthonormal
coordinates
$\tau = \sqrt{1-\frac{R_S}{r_0}} t$,
$\rho = \frac{r}{\sqrt{1-\frac{R_S}{r_0}}}$ around
$r=r_0$. With these,
equation (\ref{srgwe1}) takes the form
\begin{align}
\bigg(-\frac{1}{c^2}\frac{\d^2}{\d\tau^2}+\frac{\d^2}{\d\rho^2}+
\frac{2-3\frac{R_S}{r_0}+\frac{R_S^2}{r_0^2}}{\rho}
\frac{\d}{\d\rho}\bigg)G =& -2c\delta(\tau)\delta(\rho)
\end{align}
This equation is of the same form as equation (\ref{ddimrad}), but
this time the factor in front of the first order radial derivative
varies between 0 (in the limit $r_0\rightarrow R_S$) and 2
(in the limit $r_0\rightarrow\infty$). Except for in the latter limit,
these factors are too small to make the Green function reach
zero behind the wave front. Hence, there will be dispersion.

The right hand side of this equation is problematic, as the source term
is in the middle of the black hole. This does not matter much, though,
since we are interested in the fate of an initially sharp wave front,
and we only need the homogeneous version of the equation to find the
time evolution of such a wave. One could, for example, start with
an initial wave concentrated in a thin shell centered on the black hole
(a shell because we assume spherical symmetry here). The homogeneous
equation then tells us that this will evolve as in a $3(1-\frac{R_S}{r_0})
+\frac{R_S^2}{r_0^2}$-dimensional
spacetime, and thus disperse accordingly.

Though we considered the scalar wave equation here, one will in
general find a tail for the Green function in curved spacetime
also for the electromagnetic wave equation \citep{poisson}.

The existence of a tail of the Green function means that parts of the
field spread at speeds slower than the speed of light, and thus
move inside the light cone. In general, parts of the field
will be found to move at all speeds slower than the speed of light
\citep[pg. 11]{poisson}. When a sharp wave front enters an area of curvature
from flat spacetime, it will begin developing a trailing tail stretching
not only to the point where the curved area starts, but also beyond this,
since dispersion
also leads to part of the field moving in the opposite direction to that of the
initial wavefront. This part can propagate back to the particle and interact
with it.

There are, then, 4 ways for the particle to interact with
its own field:
\begin{itemize}
\item Locally, where the particle reacts to the field it just emitted.
\item By taking a different path through spacetime than the field, which
lets it encounter even the parts of the field that move at the speed of light.
\item By catching up to parts of the field it emitted earlier, because
the field is moving slower than $c$.
\item By having parts of the field scattered back at it.
\end{itemize}

\section{The solution of the energy paradox}
\label{solsect}
The question is: How does energy conservation come about in a situation
where a charged particle moves in the Schwarzschild spacetime, say in
the equatorial plane, under the action of gravity alone?
Assuming the particle moves in a circular orbit\footnote{If we are
to have any hope of solving the energy paradox, the orbit cannot
really be circular - the particle must spiral inwards - but far away
from the mass this effect will be small, and only makes for a negligible
correction to the circular orbit.} far away at a distance
$r$ from a mass $M$,
it will have a motion given by
\begin{align}
\mathbf r &= r\mathbf e_{\hat r} &
\mathbf v &= r\omega\mathbf e_{\hat\theta} \notag \\
\mathbf a &=
-r\omega^2\mathbf e_{\hat r} & \dot{\mathbf a} &= -r\omega^3
\mathbf e_{\hat\theta} \label{classcirc}
\end{align}
with $a\equiv |\mathbf a| = GMr^{-2}$, giving
a radiated effect of
\begin{align}
P = \frac{q^2G^2M^2}{6\pi r^4\epsilon_0c^3}\label{Pradc}
\end{align}
by the use of equation (\ref{prad}), and the particle needs to lose
energy at the same rate for energy to be conserved.

The answer to the question above
is contained in the generalized LAD-equation for curved space, equation (\ref{GLAD}).
In the present case the local term containing the Ricci curvature tensor
does not contribute since this tensor vanishes in the external Schwarzschild
spacetime. It is the non-local so-called tail-term that matters here.

Before we discuss the solution of the paradox, let us
note an interesting conceptual difference between the special relativistic
version of the equation of motion as deduced by Lorentz and Abraham on
the one hand, and the fully covariant version of it deduced by Dirac.
In the present case we are not concerned with velocity-dependent effects.
Hence we may start by looking at the non-relativistic version of the
equation,
\begin{align}
\mathbf F_{\textrm{ext}}+\alpha_q\dot{\mathbf a}=&m\mathbf a
\end{align}
The special relativistic (and Newtonian) acceleration is absolute
and arises when the particle is acted upon by an external force (or during
pre-acceleration or run-away motion). Hence, during circular motion
around the Earth under the action of the gravitational force, there
is an external gravitational force $\mathbf F_\textrm{ext}=\frac{GMm}{r^2}$
acting in the radial direction. The change of the acceleration is
directed opposite to the motion, and so the field reaction force
acts like a frictional force, decreasing the velocity of the charge, which thereby
starts spiraling inwards. The radiated energy is accounted for by a
decrease in mechanical energy.

As explained by \citet{dewitt2}, the contents of the covariant equation
(\ref{LAD}) interpreted within the general theory of relativity
is different. According to this theory, gravity is not reckoned
as a force. In Newton's theory and in special relativity we say that
a freely falling particle is acted upon by a gravitational force.
In Einstein's general theory of relativity we say that a freely
falling particle is not acted upon by any force. Such a particle has
vanishing four-acceleration. DeWitt and DeWitt then say:
\begin{quote}
When a charged particle is accelerated by means of nongravitational forces, the
electric field lines which emanate from it bend and redistribute themselves in
the vicinity of the particle (i.e., within a distance of the order of the
classical radius) in such a way as to exert, on the average, a net retarding
force, over and above the force of inertial reaction. With purely gravitational
forces, however, this is not the way things happen. The field in the immediate
vicinity of the particle tends to fall freely with the particle, and although
it suffers a local tidal distortion characteristic of an explicit occurrence of
the Riemann tensor \ldots the net retarding force due to this distortion is
zero when integrated over solid angle\footnote{
The statement that there is no local contribution to the force on the particle
for free fall is not strictly correct. It is based on a result by \citet{brehme},
which was later corrected by \citet{hobbs}. But this correction only applies in the
case of $R_{\mu\nu}\ne 0$ at the position of the particle, and so does not apply
to the case of a charged test particle in orbit around a mass, which they go on
to consider.}.
\end{quote}
Hence, the usual Abraham force plus local contributions to the field reaction
force from terms depending upon the curvature vanish for free particles.
However, DeWitt and DeWitt continue, saying:
\begin{quote}
The deviation of the particle motion from geodesic when
$F^\mu_\textrm{ext} = 0$ is caused not by the local field of the particle
but by a field which originates well outside the classical radius and
which is manifested by the nonlocal term of equation [(\ref{GLAD})].
\end{quote}
They calculated this by considering the whole electromagnetic field
(and not just the part of it in the neighborhood of the particle, as one
normally would)
using the weak field approximation and the assumption of a static gravitational
field. The result is that
the curved spacetime around a massive object
disperses and scatters the electromagnetic field. Thus, it is possible
for the field to emanate from the particle, propagate until it reaches an area
with non-negligible curvature, be scattered there, and then propagate back to
the particle. The DeWitts write
\begin{quote}
Physically [this] arises from a back-scatter process in which the Coulomb field
of the particle, as it sweeps over the ``bumps'' in space-time, receives
``jolts'' which are propagated back to the particle.
\end{quote}
This non-local interaction between the particle and its field provides a mechanism
for reducing the particle's energy and balancing the energy budget. But
as opposed to the radiation reaction for nongravitational forces, this
force is determined not by the particle's current motion, but by
its motion at some earlier time, and so there is no reason to
expect these to match, or even resemble each other.

This non-locality also makes the equation of motion a time delay
differential equation, a difficult class of differential equations
that often lack analytical solutions. However, by going to the non-relativistic
limit, DeWitt and DeWitt succeed in evaluating the extended field reaction
for a point charge in free fall near some point mass, and near some arbitrary
mass distribution, the result being
\begin{align}
\mathbf F =& \mathbf F_C + \mathbf F_{NC} \notag \\
\mathbf F_C =& q^2GM\frac{\mathbf r}{r^4} \notag \\
\mathbf F_{NC} =& -\frac 2 3 q^2 GM \left(\frac{\mathbf 1}{r^3}
- 3\frac{\mathbf r \mathbf r}{r^5}\right) \cdot \dot {\mathbf r} \label{Fpoint}
\end{align}
where $M$ is the mass of the source of the gravitational field and $q$ is
the charge and $\mathbf r$
is the separation 3-vector between the charge and the mass as measured in the
rest frame of the mass. This can be written
\begin{align}
\mathbf F_C =& -m\nabla\psi \notag \\
\mathbf F_{NC} =& -\frac 2 3 q^2 \dot{\mathbf r} \cdot \nabla\nabla \phi =
-\frac 2 3 q^2\frac{d}{dt}(\nabla\phi) \notag \\
\psi(\mathbf r) =& \frac 1 2 \frac{q^2}{m} G\int \frac{\rho(\mathbf r')}{
(\mathbf r-\mathbf r')^2}d^3r' \notag \\
\phi(\mathbf r) =& -G\int \frac{\rho(\mathbf r')}{|\mathbf r-\mathbf r'|}d^3r' \label{cncforce}
\end{align}
where $\rho$ is the mass density and $m\phi$ is the normal gravitational field.
The mathematical expressions for $\mathbf F_{NC}$ are explained in Appendix
\ref{notationapp}.
DeWitt and DeWitt write
\begin{quote}
$\mathbf F_{NC}$ is the nonconservative force which gives rise to radiation
damping. Owing to its dependence on velocity it is small in magnitude compared
to the force $\mathbf F_C$ which is conservative. \ldots It [the conservative
part] arises from the fact that the total mass of the particle is not
concentrated at a point but is partly distributed as electric field energy in
the space around the particle.
\end{quote}
So $\mathbf F_{NC}$ is the force that should correspond to the radiated effect
in equation (\ref{Pradc}). However, this effect was calculated under the
assumption of a circular orbit, which is not possible under the influence
of a nonconservative damping force. But if $\mathbf F_{NC}$ is very small, as it
should be, because the field has traversed a distance of $2r$ before
returning to the particle, and only a small part of the field is scattered
back in the first place, then the particle motion will be dominated by
the force of gravity, so $F_\textrm{total}\approx F_G = -m\nabla\phi$.

By inserting this into the expression (\ref{cncforce}) for $F_{NC}$,
we see that the nonconservative force, which with its dependence on
velocity and the local shape of $\phi$ is quite different from the
Lorentz-Abraham force $\mathbf F_{LA} = \alpha_q \dot{\mathbf a}$, still
reduces to it in the non-relativistic limit. The work this force does on the
particle is given by $P = \mathbf F_{NC}\cdot \mathbf v$, and using
equation (\ref{classcirc}), we find
\begin{align}
P =& \alpha_q \dot{\mathbf a}\cdot\mathbf v =
 -\alpha_q r^2\omega^4 = -\alpha_q a^2 =\frac{q^2G^2M^2}{6\pi r^4\epsilon_0c^3}
\end{align}
which is of the same size but opposite magnitude as the radiated effect
in equation (\ref{Pradc}), meaning that the particle loses energy at the
same rate as it is radiated away, and energy is thus balanced.

This applies not only for the circular orbit considered here; DeWitt
and DeWitt showed (still using the assumption that the motion is dominated
by the effects of gravity) that the average energy loss rate during
any closed orbit in free fall matches that given by the Larmor formula.
This follows from the fact that the nonlocal, nonconservative,
backscattering-based force $\mathbf F_{NC}$ reduces to normal Lorentz-Abraham
force $\mathbf F_{LA}$ in the non-relativistic limit. This, then, is what balances the
energy budget, and prevents an electron in orbit around the Earth from
being a perpetual source of energy.

\section{Acknowledgments}
We would like to thank Jon Magne Leinaas for useful discussion concerning
the electromagnetic perpetuum mobile paradox discussed here.

\appendix
\section{}
\label{notationapp}
Given two vectors $A^\mu$ and $B^\mu$, two obvious products can be
formed: the inner product $P = A^\mu B_\mu$, and the outer product
$Q^{\mu\nu} = A^\mu B^\nu$. DeWitt and DeWitt indicate the former
with the $\cdot$ operation, and the latter by adjoining two vectors
without any intervening notation. Thus, the $\mathbf r \mathbf r$ seen
in equation (\ref{Fpoint}) is
\begin{align}
\mathbf r\mathbf r = r_i r_j = \left(\begin{matrix}
r_1 r_1 & r_1 r_2 & r_1 r_3 \\
r_2 r_1 & r_2 r_2 & r_2 r_3 \\
r_3 r_1 & r_3 r_2 & r_3 r_3
\end{matrix}\right)
\end{align}
and $(\mathbf r\mathbf r)\cdot \dot {\mathbf r} = \mathbf r (\mathbf r\cdot
\dot{\mathbf r})$. The matrices in equation (\ref{Fpoint}) simply serve
to let one write the $\dot{\mathbf r}$ as a single factor outside
the parentheses; otherwise, it could be written without them as
\begin{align}
\mathbf F_{NC} =& -\frac 2 3 q^2 GM \left(\frac{\dot{\mathbf r}}{r^3}
- 3\mathbf r\frac{\mathbf r \cdot \dot{\mathbf r}}{r^5}\right)
\end{align}

The above also applies to the vector operator $\nabla$, which is
$(\d_x,\d_y,\d_z)$ in Cartesian coordinates. The inner product
formed from $\nabla$ is $\nabla^2 \equiv \nabla\cdot\nabla
= \d_x^2+\d_y^2+\d_z^2$, but one can also form the outer product
\begin{align}
\nabla\nabla \equiv \left(\begin{matrix}
\d_x\d_x & \d_x\d_y & \d_x\d_z \\
\d_y\d_x & \d_y\d_y & \d_y\d_z \\
\d_z\d_x & \d_z\d_y & \d_z\d_z
\end{matrix}\right)
\end{align}
While $\nabla^2\phi = 0$ in areas where the mass density $\rho=0$,
the same does not apply to $\nabla\nabla\phi$; otherwise, $F_{NC}$
would be 0 at the particle's position, and we would have no extended
field reaction.

Note also that the fact that the gravitational field is static
has been used in the transition $\mathbf v \cdot \nabla\nabla\phi
= \frac{d}{dt}\nabla\phi$. In general, the total time derivative
of some field $\xi$ as experienced by a particle moving with velocity
$\mathbf v$ is given by
\begin{align}
\frac{d}{dt}\xi =& \frac{\d}{\d t}\xi + \mathbf v\cdot\nabla \xi
\end{align}
but the partial derivative of $\xi = \nabla\phi$ with respect to time
is 0 here.

\bibliographystyle{unsrtnat}
\bibliography{refs.bib}

\end{document}